# Possible Formation of QGP-droplets in Proton-Proton Collisions at the CERN Large Hadron Collider

RAGHUNATH SAHOO*
DISCIPLINE OF PHYSICS, SCHOOL OF BASIC SCIENCES,
INDIAN INSTITUTE OF TECHNOLOGY INDORE, SIMROL, KHANDWA ROAD, INDORE-453552, INDIA

## ABSTRACT

Proton-proton (pp) collisions have been traditionally used as a baseline measurement in the search for a deconfined state of matter in heavy-ion collisions at ultra-relativistic energies. The unprecedented collision energies that are available at the Large Hadron Collider (LHC) at the European Laboratory for Nuclear Research (CERN) have illuminated new challenges in understanding the possible formation of droplets of this deconfined matter of partonic degrees of freedom in hadronic collisions, especially in high-multiplicity events. Enhancement of multi-strange particles compared to pions [1], degree of collectivity [2,3], comparable freeze-out temperature with heavy-ion collisions [1], observation of a long-range ridge-like structure for high-multiplicity events [3,4] are some of the experimental observations in this direction. In this article, we discuss some of the experimental observables and outline new theoretical directions to understand the possibilities of exploring the formation of QGP-droplets in pp collisions at the LHC.

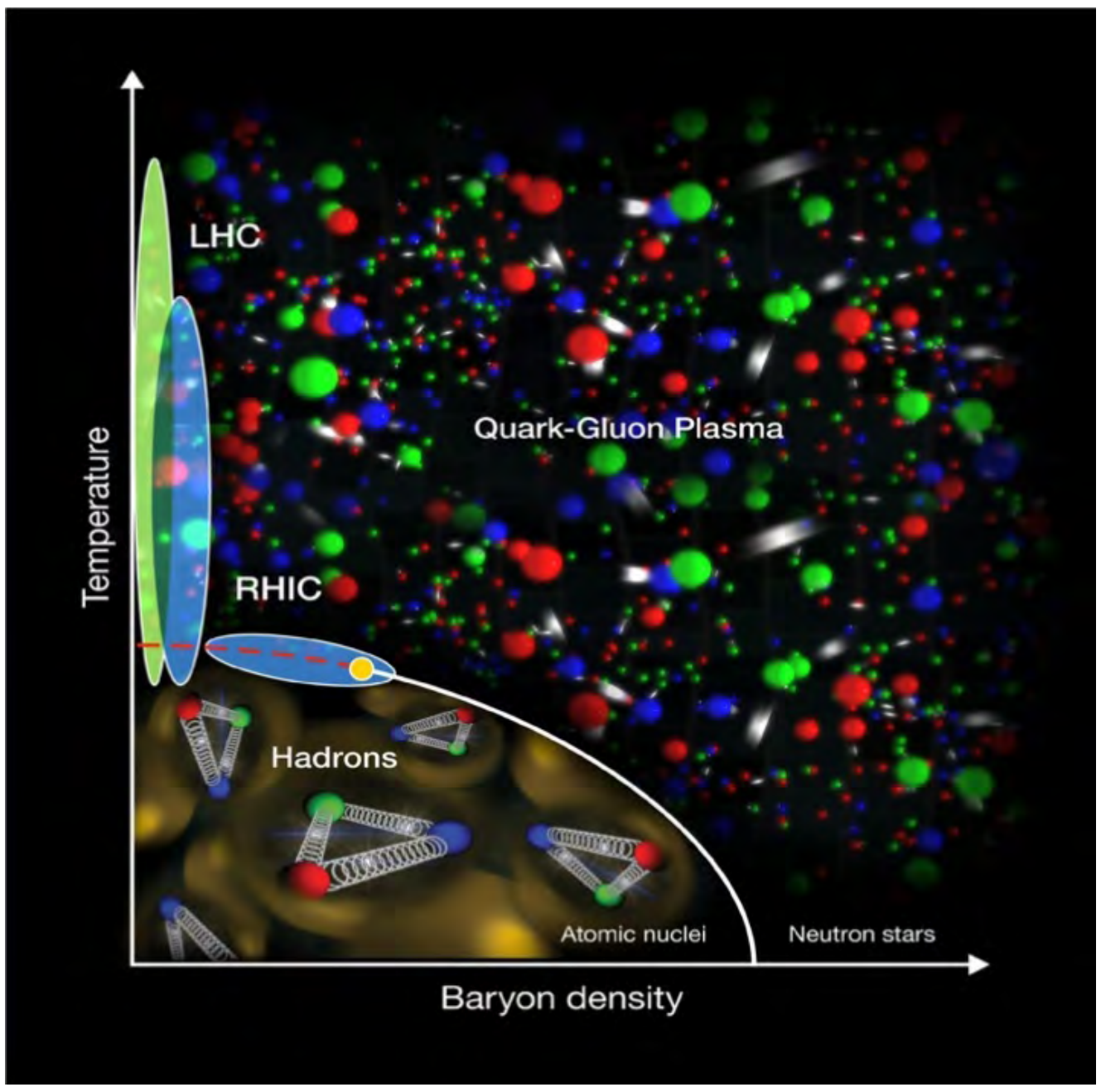

**Fig. 1:** A schematic of the QCD phase diagram showing a possible critical point. Figure adopted from [32].

## INTRODUCTION

The quest to understand the fundamental structure of matter has driven us to probe the sub-atomic universe. With the advent of high energy accelerators, through collisions of hadrons or heavy-nuclei, it has been possible to create a state of matter, usually described by the partonic degrees of freedom. Partons (quarks and gluons) being the quanta of the system, designate the created matter to be primordial and should be governed by QCD, i.e., quantum chromodynamics, the theory of strong interaction, which manifests itself at a low-length scale or high momentum transfer. Although QCD's fundamental nature suggests that because of its peculiar property of *"asymptotic freedom and infrared slavery"* [5] free partons can't be observed in the laboratory, there have been attempts to create free quarks, at least a plasma of partons. In these efforts, in nuclear collisions at the Relativistic Heavy-Ion Collider (RHIC), Brookhaven National Laboratory, USA and at the LHC at CERN, scientists have been able to form a new state of matter called quark-gluon plasma (QGP). QGP is defined as a (locally) thermally equilibrated system where the partons are deconfined from the hadrons so that the color degrees of freedom become manifest in nuclear rather than the nucleonic volume [6]. This can be achieved either by heating the nuclei to ultra-high temperatures (hundreds of MeVs, 1 MeV = $1.16 \times 10^{10}$ Kelvin), which is done at RHIC and LHC energies; or by compression of nuclei so as to diffuse the hadronic boundaries. The latter is expected to take place at the Facility for Antiproton and Ion Research (FAIR) at GSI, Germany and the Nuclotron-based Ion Collider fAcility (NICA), Joint Institute for Nuclear Research (JINR), Russia. In Fig. 1, we show a typical QCD phase diagram (schematic), where the temperature is plotted as a function of the baryon density of the system. The deconfined phase of quarks and gluons is separated by a first order phase transition line from the confined hadronic matter, which ends with a possible critical end-point (CP). Thereafter, one observes a cross-over transition in the RHIC and LHC energy regimes. The exploration of this QCD phase diagram and the search for the CP has been a frontier of high energy nuclear research for decades. In this phase diagram, a high temperature and low baryon density corresponds to an early universe scenario that might have existed billions of years ago, whereas the low temperature and high baryon density regime corresponds to different astrophysical objects like neutron stars. It is believed that QGP might have existed at the early universe or exists at the core of neutron stars.

*E-mail address: Raghunath.Sahoo@cern.ch

The critical energy density and temperature required for such a deconfinement transition is $\varepsilon_c$ = 1 GeV/fm$^3$ and $T_c \approx$ 150-170 MeV, respectively. This is the estimation given from lattice QCD calculations [7]. It should be noted here that the energy density of normal nuclear matter is $\rho_N$ = 0.17 GeV/fm$^3$ and the energy density of a nucleon is $\rho_p$ = 0.5 GeV/fm$^3$ (proton charge radius ~ 0.84 fm). To have a physical idea of the critical temperature, it is almost $10^5$ times the core of the Sun's temperature. For a hadron gas, the degrees of freedom are mostly three (pion isospin), whereas for QGP (2-flavors: u, d quarks) the degrees of freedom become 37: orders of magnitude higher [8].

## THE SPACE-TIME EVOLUTION

The space-time evolutions in hadronic and heavy-ion collisions (are shown in Fig.2) are complex phenomena involving various degrees of freedoms at different space-time coordinates. Due to this complexity it is very difficult to describe the whole system's evolution with a single theory. After the collisions of heavy-ions, the system goes through a pre-equilibrium phase followed by a deconfined QGP medium and a possible mixed phase (which should show first order phase transition signatures). Hadronization occurs after this, forming composite hadrons from the primordial partonic matter. This point is designated by the critical temperature ($T_c$), where a possible phase transition from QGP to hadron gas occurs. The chemical composition of the system is frozen thereafter (inelastic collisions happen to cease), making the particle ratios fixed with time. This is characterized by the chemical freeze-out temperature ($T_{ch}$) and the baryochemical potential ($\mu_B$) of the system. A statistical hadron gas model works fine at this point, which ignores the interactions treating the system as an ideal gas. The final state particles then fly toward the detectors and the mean free path of the system becomes higher than the system size, making the particles collide, infrequently, with each other. At this point, the transverse momentum distribution of the system is fixed with time (elastic collisions stop). This is called kinetic freeze-out. For an equilibrium distribution, usually one fits a Boltzmann distribution to describe the particle spectra (with invariant yield as a function of transverse momentum) and the inverse slope parameter is the *effective temperature* of the system, when a particular particle comes out of the system. Fitting the Boltzmann-Gibbs blast-wave model to the low-$p_T$ part of the spectra (usually less than 3 GeV/c), one obtains the collective radial flow velocity of the particle. There are two different schools of thoughts; in one scenario a single freeze-out - technically a simultaneous fit - is performed to all the identified particle spectra to obtain a common radial flow velocity and thermal temperature; the other scenario is a differential (mass-dependent) freeze-out scenario, which shows massive particles coming out of the system earlier in time with smaller radial flow velocities, which is hydrodynamic behavior. In addition, there are also scenarios like strange and non-strange particles freezing out separately. The full $p_T$-spectrum is a composition of hard and soft domains of physics, the hard being described by pQCD and the soft physics is non-perturbative, and usually described by statistical models.

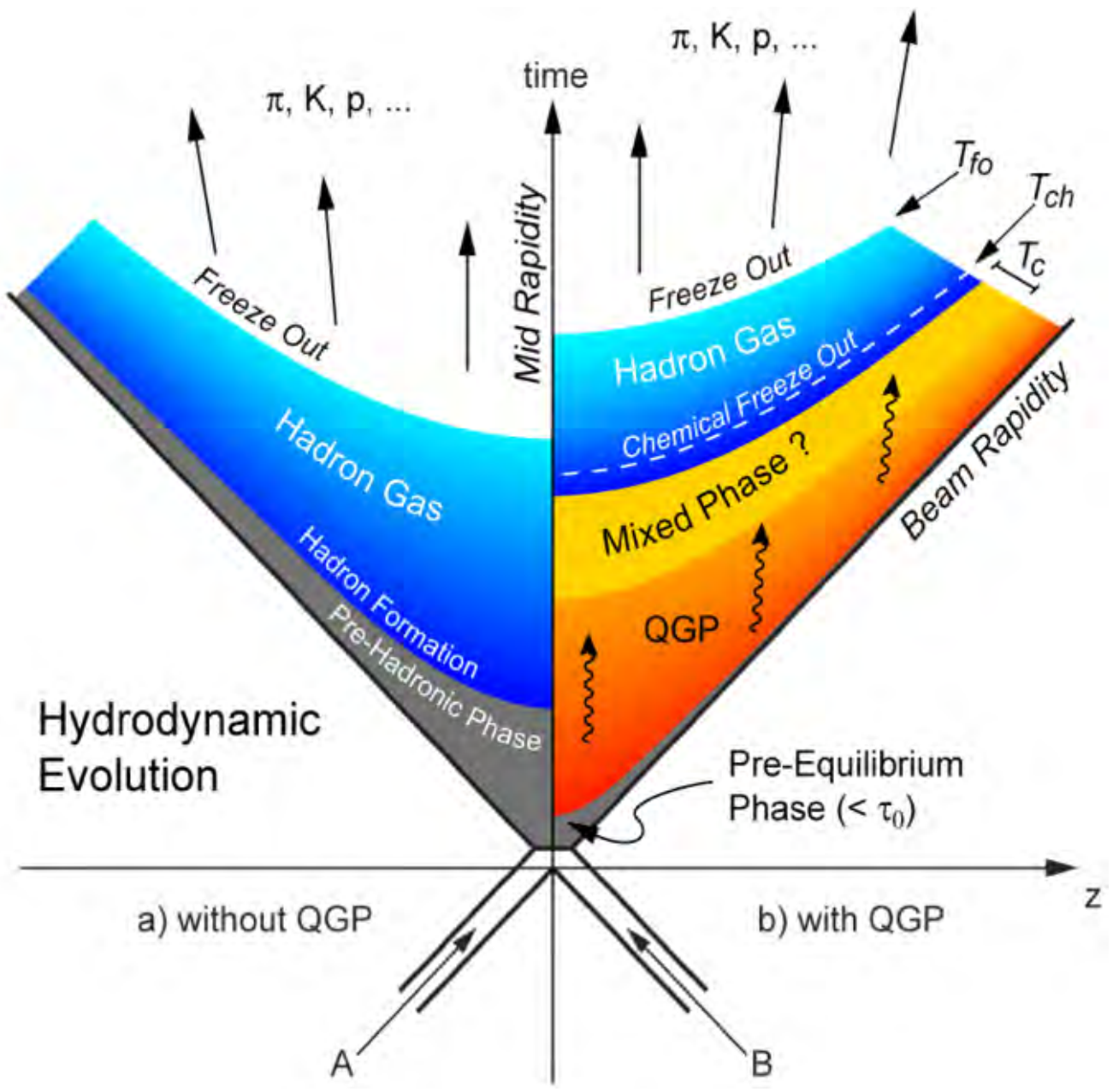

**Fig. 2:** A possible space-time Minkowski diagram in hadronic collisions, compared with heavy-ion collisions. Figure adopted from [33].

However, the space-time evolution of the system in hadronic (pp) collisions are the least explored area and has lately become a new domain of physics. As seen from Fig. 2, it is believed that in hadronic collisions the system evolves, through a pre-hadronic phase, hadron formation (hadron gas) and then freeze-out occurs. However, scientists have recently adopted the idea of a similar space-time picture as that of a heavy-ion collision, at least at the LHC energies, thereby consolidating the avenues for direction in hadronic physics. We shall discuss these new aspects of LHC hadronic collisions in the subsequent sections.

## SOME SIGNATURES OF QGP

As the lifetime of QGP is of an order of $10^{-23}$ of a second, the signatures are indirect in nature. Here we list a few of the important signatures of QGP; for completeness, one can review any standard book on QGP [9,10]. It should also be noted here that, with time, as we go into the new horizons of research in energy frontiers, some of the concepts involved in the signatures of QGP become modified and new discoveries emerge carrying new signatures.

### a) Strangeness enhancement

Colliding objects like protons or heavy-ions are composed of $u$ and $d$ quarks, as the constituents of the nucleons. If we are able to form composite objects, e.g., hadrons with a strange quark as the constituents, these objects must be formed as a part of a possible partonic medium through a process called hadronization or in particles producing inelastic interactions. However, normally we compare the experimental observations on strangeness enhancement with a hadron gas scenario at a particular collision energy to have a proper conclusion. In equilibrated QGP, as the temperature of the system is higher than the mass of a strange quark ($m_s \gg m_{u,d}$), strange quarks and anti-quarks can be abundantly produced through several processes leading to strangeness enhancement. These processes are- flavor creation ($qq \rightarrow s\bar{s}$, $gg \rightarrow s\bar{s}$), gluon splitting ($g \rightarrow s\bar{s}$) and flavor excitation ($gs \rightarrow gs$, $qs \rightarrow qs$). An observation of enhanced multi-strange particles in comparison to the pion yields in the final state is called strangeness enhancement and is considered to be a signature of QGP. Strangeness enhancement has been experimentally observed from CERN SPS [11], RHIC [6] and LHC [12] energies in heavy-ion collisions.

### b) Suppression of J/ψ

J/ψ is a bound state of charm and anti-charm quarks. In a QGP medium, in the presence of various other quarks, anti-quarks and gluons, a charm quark may not come in the vicinity of its counterpart to form a bound state; this is called color Debye screening. This is analogous to the screening of electric charges in an electromagnetic plasma (in QED), where this phenomenon is called QED Debye screening. In addition, in a QGP phase as a deconfined medium at very high temperatures, the string tension between charm and anti-charm quarks vanishes. In this situation, in a combined effect of both of these, the yield of J/ψ is suppressed [13]. In contrast, the yields of open charms ($D^0$, $D^{\pm}$, $D_s$ etc.) are enhanced because the orphan charm quarks can combine with other light-flavors found in the proximity. The suppression of J/ψ serves as a signature of QGP. However, it could be possible that when the collision energy is high enough to produce plenty of charm/anti-charm quarks in the system, there could be competition between suppression and regeneration/recombination. Higher states of quarkonia can dissociate/decay to form more J/ψ. This is the reason the degree of suppression of J/ψ in the LHC heavy-ion collisions is lesser compared to RHIC heavy-ion collisions [14].

### c) Suppression of high-$p_T$ particles (jet quenching)

In hadronic or nuclear collisions at relativistic energies, high-$p_T$ particles are produced from the initial partonic interactions and serve as an indirect probe of a possible deconfined medium. From the typical partonic scatterings ($g + g \rightarrow g + g$, $q + q \rightarrow q + q$), quarks and gluons with very high transverse momenta are produced, which fragment to create a multitude of correlated particles, usually, in a conical volume called "jets". Hard partons are produced early in time (formation time $\tau \sim 1/p_T$, $p_T$ being the component of momentum in the direction transverse to the beam) and hence they are useful in probing the early stages of the collision. These high-$p_T$ partons/jets lose energy as they traverse through the high-density medium. This energy loss is dependent on the path length. In other words, a parton/jet produced nearer to the periphery of the fireball, if it escapes directly, will lose less energy compared to a parton/jet that was produced inside the bulk of the medium. One constructs a $p_T$-dependent observable called a nuclear modification factor, $R_{AA}$, which is defined as:

$$R_{AA}(p_T) = \frac{1}{\langle N_{coll} \rangle} \frac{Yield_{AA}}{Yield_{pp}}$$

Here $\langle N_{coll} \rangle$ is the mean number of binary nucleon-nucleon collisions occurring in a single nucleus-nucleus collision. This is obtained through Glauber model estimations [15]. If the nucleus-nucleus collision were a simple superposition of nucleon-nucleon collisions (no medium effect), then $R_{AA}$ would be unity. It is seen by the CMS experiment at the LHC that for the mediating particles of weak interactions, $W^{\pm}$, $Z^0$ [16] and also for direct photons [17], there is no nuclear suppression, as was expected. However, for high-$p_T$ (identified) charged particles, the degree of suppression (~ 4-5 times) is much higher, $R_{AA} < 1$ in Au-Au or Pb-Pb collisions, compared to dAu or pPb collisions, where QGP is not expected to be formed. The observation of no suppression in dAu or pPb collisions is an indication of the no cold nuclear matter (CNM) effect at high-$p_T$. This observation translates to the fact that hard-scattered partons lose energy while travelling through the hot and dense medium, and the suppression is understood to be a final state effect.

Jet quenching was nicely seen by the STAR experiment at RHIC through the dihadron angular correlation [18]. Considering a high-$p_T$ particle as the trigger particle ($4 < p_T^{trig} < 6$ GeV/c), the azimuthal angle between the trigger particle and the associated particles ($2 < p_T^{assoc} < p_T^{trig}$ GeV/c) was studied for pp, dAu and Au-Au collisions. The suppression of an away-side jet ($\Delta\phi \sim \pi$) was observed in Au-Au collisions, which was absent in pp and dAu collisions. This evidenced the formation of a hot and dense medium in heavy-ion collisions making the away-side partons lose their energy while traversing through the medium and thereby producing low-$p_T$ particles.

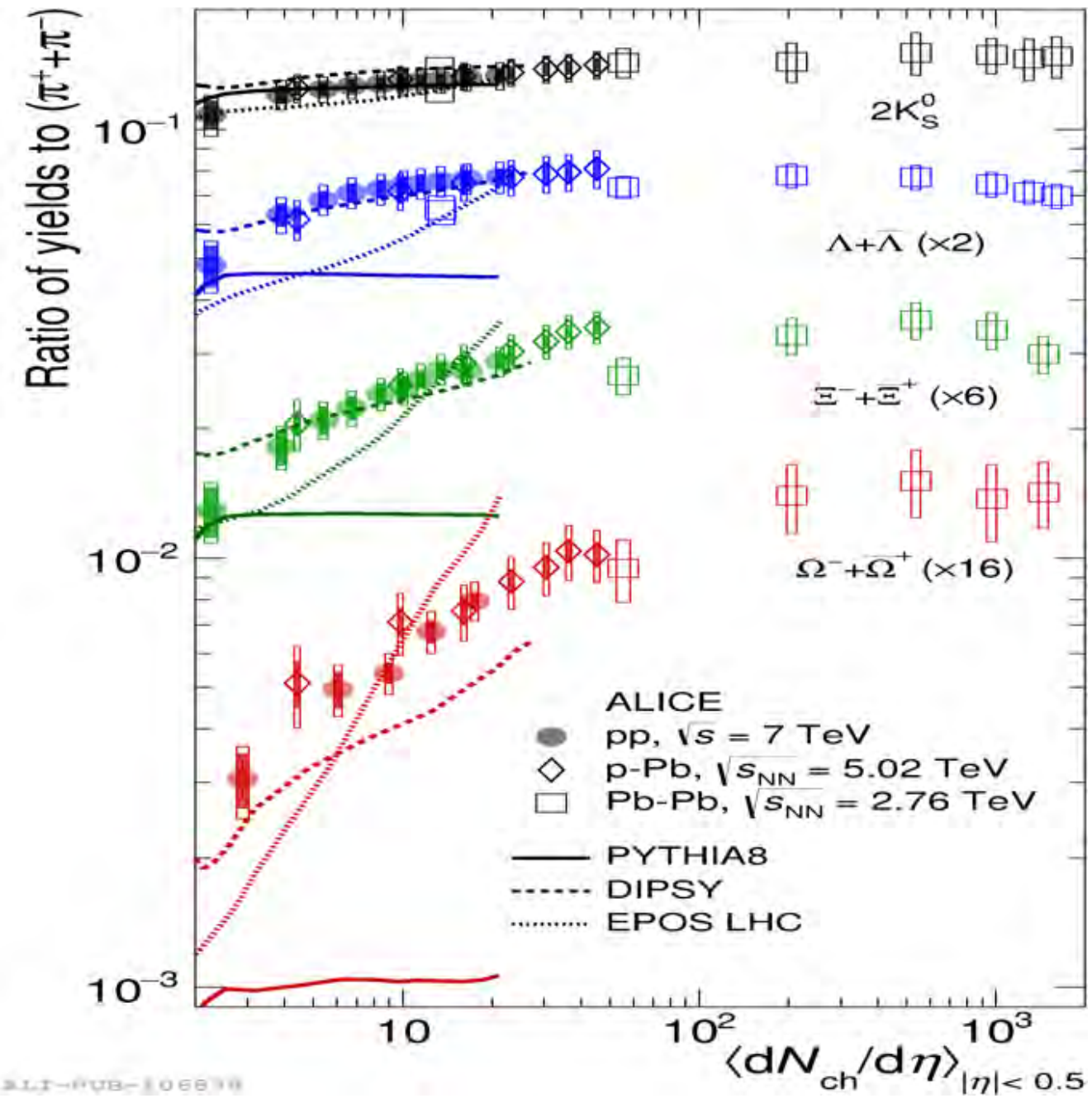

**Fig. 3:** Transverse momentum integrated yield ratios of strange and multi-strange hadrons to pions as a function of mid-rapidity charged particle density. The measurements in pp collisions at $\sqrt{s}$ = 7 TeV are compared with the results from p-Pb and Pb-Pb collisions and with theoretical MC models. This shows heavy-ion-like strangeness enhancement in high-multiplicity pp collisions. Figure adopted from [1].

## NEW OBSERVATIONS IN PP COLLISIONS

### a) Enhancement of Multi-strange Particles

The ALICE experiment at the LHC has studied the enhanced production of multi-strange particles in high-multiplicity pp collisions at the midrapidity (rapidity $y = (1/2) \ln (E+p_zc)/(E-p_zc)$, where E is the energy of a particle of mass, $m$ and $p_z$ is the component of momentum along the beam direction, c is the speed of light in vacuum) at a center-of-mass collision energy of 7 TeV ($\sqrt{s}$ = 7 TeV) [1]. Taking the ratios of $p_T$-integrated yields of multi-strange particles with respect to pions (lowest mass bosons, which dominate the final state in a muti-particle production process) as a measure of strangeness production, it is observed (Fig. 3) that the high-multiplicity pp collisions show similar values like heavy-ion (Pb-Pb) collisions, where a QGP is formed [1]. This is an excellent observation at the LHC energies, paving a way to have a deeper look into high-multiplicity events, their origin from a microscopic picture, observables effects and at the end to conclude if QGP-droplets are produced in such collisions. We shall take few more observations in pp high-multiplicity collisions to establish the path forward in this direction of research.

**b) Transverse momentum Spectra, Collectivity**

The only function that well describes the $p_T$-spectra of identified particles in pp collisions at the LHC energies is the Tsallis-Levy. Although this method lacks with a microscopic description of the system, because of its spectral description, it has been widely used both by experimentalists and theoreticians to extract the particle yields and also to study various thermodynamic features for systems like pp, which are assumed to be away from equilibrium. However, to extract the collective radial flow velocity of the system, one takes a low-$p_T$ part of the spectra (usually up to $p_T \sim 3$ GeV/c) and makes a simultaneous fit of blast-wave model to identified particle spectra. By doing this, one obtains the average radial flow velocity and also the thermal temperature at the kinetic freeze-out. For pp collisions at 7 TeV collision energy, such an analysis on multi-strange particle spectra gives a freeze-out temperature, $T_{fo} = 163 \pm 10$ MeV and radial flow velocity, $\langle\beta\rangle = 0.49 \pm 0.02$ [1]. This temperature remarkably falls within the range of temperature required for a deconfinement transition as per the lattice QCD estimates [7]. The high-multiplicity pp events seem to show a high degree of collectivity as well.

**c) Multiparticle Ridge-like Correlations**

Ridge is a long-range near-side structure in two particle azimuthal correlations and was first observed at RHIC in Cu-Cu [19] and Au-Au [20] collisions and later at the LHC, in Pb-Pb collisions [21]. The reason of this ridge formation in heavier collision systems like heavy-ions is understood to be due to hydrodynamic collective flow of strongly interacting matter, which undergoes expansion, developing long-range (in rapidity, $|\Delta\eta| \approx 4$) correlations. The CMS experiment at the LHC, has observed [3] a same-side ($\Delta\phi \sim 0$) ridge in the two particle correlations produced in high-multiplicity pp collisions. This, along with several other observations on collectivity in small systems, has opened up a new direction in understanding particle production in small systems like pp collisions.

These are some of the novel new phenomena observed in the high-multiplicity pp collisions at the LHC energies, which warrant deeper understanding.

## EMERGENT PHENOMENA

Here we list some emergent phenomena that we believe could enlighten researchers in the field. It should be mentioned here that the list may not be an exhaustive one.

a) The strangeness enhancement in high-multiplicity pp collisions is explained by theoretical models like DIPSY, where the interactions between gluonic strings are allowed to form color ropes and through the mechanism of *"rope hadronization"*, strangeness is produced.

b) The flow-like features in such high-multiplicity events are explained by *"color reconnection"* mechanisms in the PYTHIA event generator.

c) The observation of long-range (rapidity) azimuthal correlations in pp collisions, at the near-side ridge, indicates possible hydrodynamic collective behavior of strongly interacting medium like that observed in heavy-ion collisions.

d) *"Multipartonic interactions (MPI)"* seem to explain the production of various light and heavy-flavor particles [15].

e) The theory of *"non-extensivity"*, by explaining the particle spectra up to very high-$p_T$ (Tsallis-Levy distribution), has created a new direction in hadronic collisions- systems away from equilibrium. System thermodynamics, the degree of equilibration as a function of final state particle multiplicity, the differential freeze-out scenario, connecting particle production in pp collisions to that of heavy-ion collisions using the Boltzmann transport equation in relaxation time approximation, etc. are some of the new features of high-energy research [22, 23].

f) *Event shape engineering*, separating jetty and isotropic events in LHC pp collisions through transverse spherocity, has become a new tool for exploring particle production mechanisms [24, 25, 26]. Jetty events are usually governed by high-$p_T$ phenomena and hence are described by pQCD (perturbative quantum chromodynamics), whereas the isotropic events are soft physics dominated. The particle ratios mean transverse momentum, event thermodynamics, etc. could be studied as a function of transverse spherocity.

g) The particle yields in the final state is seen to be a linear scaling function of the charged particle multiplicity, irrespective of collision species and collision energy [27, 28]. Final state multiplicity driving particle production is a new and remarkable observation at the LHC energies. This, in fact, is an area that needs deeper understanding. When the collision energy or the initial energy density should be the driving force for the formation of a system and its further time evolution, the particle yields showing such a scaling behavior with final state multiplicity is thought provoking and requires minute understanding.

h) A final state charged particle multiplicity, $N_{ch} \geq 20$ is found to be a thermodynamic limit where different statistical ensembles are observed to give the same results [29]. This number is also important in the sense of a microscopic picture of the system as we have also seen it be a threshold for MPI to play a major role in the particle production [30].

## SUMMARY AND OUTLOOK

Although it was initially believed that only central heavy-ion collisions could create a deconfined matter of QGP (but not all central collisions)[31], we are now in a domain of collision energies, where we need to digest the fact that LHC hadronic collisions create a matter that has QGP-like signatures. Going one step forward and strongly assuming that pp collisions at these energies need a different theoretical and experimental treatment, as opposed to considering them as a baseline measurement, we must look for possible QGP-droplets in pp collisions - creating a new dimension in high energy nuclear physics. Other QGP signatures seen in heavy-ion collisions are yet to be verified in high-multiplicity pp collisions. Some of these signatures are discussed in this article. Although the list of signatures is not at all exhaustive, we hope that this article has been helpful to establish a foundation of understanding for budding researchers in this field, while showing them a new horizon of exploration in the dreamland of high-energy nuclear physics.

**Acknowledgements:** The research work of the author is jointly supported by the Mega-science projects of Department of Atomic Energy and Department of Science & Technology, Government of India through the project number SR/MF/PS-01/2014-IITI(G). The author would like to recognize stimulating discussions with Dr. Jan-e Alam and Dr. Tapan K. Nayak.

### *References*

[1] J. Adam et al. [ALICE Collaboration], Nature Phys. 13, 535 (2017).
[2] A. Khuntia, H. Sharma, S.K. Tiwari, R. Sahoo and J. Cleymans, Eur. Phys. J. A 55, 3 (2019).
[3] V. Khachatryan et al. [CMS Collaboration], Phys. Letts. B 765, 193 (2017), JHEP 1009, 091 (2010).
[4] J.D. Bjorken, S.J. Brodsky and A.S. Goldhaber, Phys. Letts. B 726, 344 (2013).
[5] W. Greiner, S. Schramm and E. Stein, Quantum Chromodynamics, Springer (2006).
[6] ] J. Adam et al. [STAR Collaboration], Nucl. Phys.A 757, 102 (2005).
[7] S. Borsanyi et al. J. High Energ. Phys. 2010, 77 (2010); S. Borsanyi et al. Phys. Lett. B 730, 99 (2014).
[8] M. Kleimant, R. Sahoo, T. Shuster, and R. Stock, Lecture Notes in Physics, 785, 23 (2010).
[9] C-Y. Wong, Introduction to High-Energy Heavy-Ion Collisions, World Scientific, Singapore (1994).
[10] R. Vogt, Ultrarelativistic Heavy-Ion Collisions, Elsevier (2007).
[11] F. Antinori et al. [WA97 Collaboration] Nucl.Phys. A661, 130 (1999).
[12] B.B. Abelev et al. [ALICE Collaboration], Phys. Lett. B728, 216 (2014).
[13] T. Matsui and H. Satz, Phys. Lett. B178, 416 (1986).
[14] B.B. Abelev et al. [ALICE Collaboration], Phys. Lett. B734, 314 (2014).
[15] M. L. Miller, K. Reygers, S.J. Sanders, and P. Steinberg, Ann. Rev. Nucl. Part. Sci. 57, 205 (2007).
[16] S. Chatrchyan et al. [CMS Collaboration] JHEP 1503, 022 (2015).
[17] S.S. Adler et al. [PHENIX Collaboration], Phys. Rev. C 75, 024909 (2007).
[18] J. Adam et al. [STAR Collaboration], Phys. Rev. Lett. 91, 072304 (2003).
[19] B. Alver et al. [PHOBOS Collaboration], Phys. Rev. C 81, 024904 (2010).
[20] B. Abelev et al. [STAR Collaboration], Phys. Rev. C 80, 064912 (2009), B. Alver et al. [PHOBOS Collaboration], Phys. Rev. Lett. 104, 062301 (2010).
[21] S. Chatrchyan et al. [CMS Collaboration] JHEP 02, 088 (2014).
[22] S. Tripathy, A. Khuntia, S.K. Tiwari, and R. Sahoo, Eur. Phys. J. A 53, 99 (2017).
[23] S. Tripathy, S.K. Tiwari, M. Younus, and R. Sahoo, Eur. Phys. J. A 54, 38 (2018).
[24] A. Khuntia, S. Tripathy, A. Bisht, and R. Sahoo: arXiv:1811.04213
[25] S. Tripathy, A. Bisht, R. Sahoo, A. Khuntia, and Malavika P.S., arXiv: 1905.07418
[26] R. Rath, A. Khuntia, S. Tripathy, and R. Sahoo, arXiv:1906.04047
[27] Nucl. Phy. A, Proceedings of 27th International Conference on Ultrelativitic Nucleus-Nucleus Collisions: Quark Matter-2018.
[28] S. Tripathy [ALICE Collaboration], arXiv:1907.00842.
[29] N. Sharma, J. Cleymans, B. Hippolyte and M. Paradza, Phys. Rev. C 99, 044914 (2019).
[30] D. Thakur, S. De, R. Sahoo and S. Dansana, Phys. Rev. D97, 094002 (2018).
[31] H. Heiselberg, Phys. Rept. 351, 161 (2001).
[32] https://www.bnl.gov/newsroom/news.php?a=24281
[33] https://particlesandfriends.wordpress.com/2016/10/14/evolution-of-collisions-and-qgp/

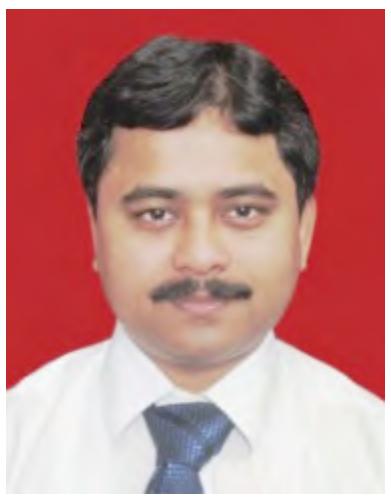

**Raghunath Sahoo** is an associate professor at the Indian Institute of Technology Indore (IIT), India. After receiving his PhD from the Institute of Physics, Bhubaneswar, he worked at Subatech, Nantes, France and the Istituto Nazionale di Fisica Nucleare (INFN) -Padova, Italy for his postdoctoral research, on the French National Center for Scientific Research (CNRS) and INFN fellowships, respectively, before joining IIT Indore in 2010. His fields of research are the experimental high-energy nuclear physics-search for quark-gluon plasma (QGP), exploration of the QCD phase diagram and phenomenological studies of matter created in hadronic and nuclear collisions at the RHIC and LHC energies. He is a member of ALICE, CERN, Geneva and the Compressed Baryonic Matter CBM experiment at GSI, Germany. An author of more than 380 experimental and around 40 international journal publications in QGP phenomenology, Dr. Sahoo is a leading researcher in the field of high-energy nuclear physics in India.